\documentclass[11pt]{article}
\usepackage{graphics}

\begin{document}
\title{A new uncertainty principle}
\author{C. Y. Chen\\
Department of Physics, Beijing University of Aeronautics\\
and Astronautics, Beijing, 100083, P.R.China}
\thanks{Email: cychen@buaa.edu.cn}
\maketitle

\vskip 50pt

\begin{abstract}
By examining two counterexamples to the existing theory, it is
shown, with mathematical rigor, that as far as scattered particles
are concerned the true distribution function is in principle not
determinable (indeterminacy principle or uncertainty principle)
while the average distribution function over each predetermined
finite velocity solid-angle element can be calculated.
\end{abstract}

\section{Introduction}

The distribution function, playing the most fundamental and most
central role in classical statistical mechanics, is defined by the
limiting process:
\begin{equation}\label{definition}
f({\bf r},{\bf v},t)=\lim\limits_{\Delta{\bf r}\to 0, \Delta{\bf
v}\to 0}\frac{\Delta N} {\Delta{\bf r}\Delta{\bf v}}
,\end{equation}
 where $\Delta {\bf r} =\Delta x\Delta y\Delta z$
is a position volume element, $\Delta {\bf v} =\Delta v_x\Delta
v_y\Delta v_z$ a velocity volume element and $\Delta N$ the number
of the particles found in the phase-volume element $\Delta{\bf r}
\Delta{\bf v}$. This concept has served us since the start of
statistical mechanics and no serious challenge was put forward
against it.

However, examining the definition (\ref{definition}) with
mathematic rigor leads us to stimulating ideas. For instance, if,
for whatever reason, some dimensions of $\Delta{\bf r} \Delta {\bf
v}$ have to be kept finite (not infinitesimal), we have no choice
but to regard them as the uncertainty ranges over which $f({\bf
r},{\bf v}, t)$ expressed by (\ref{definition}) is not truly
determinable. To exclude such uncertainty, we need to show that
the ratio in (\ref{definition}) will tend to a definite value no
matter in what way $\Delta\bf r$ and $\Delta {\bf v}$ approach
zero. This is by no means an insignificant or dispensable task in
view of the fact that many intriguing examples in math have
manifested that a multi-variable function should be considered to
have no limit wherever it has path-dependent limits\cite{courant}.
(Also note that taking partial derivatives at such places becomes
unjustified and improper.)

Being exposed to a variety of thought experiments and doing
thorough examinations on them\cite{chen1,chen2}, we are now
convinced that there indeed exist inherent constraints to limit
how $\Delta\bf r$ and $\Delta {\bf v}$ tend to zero; or, in other
words, there indeed exists a certain type of uncertainty
principle. The situation, to some extent, resembles what happens
to the uncertainty principle of quantum mechanics, which was in
history, still is in usual textbooks, obtained from analyzing
experimental facts.

Since the studies of distribution function were strongly
influenced by Boltzmann's initial approach on the Boltzmann gas
(by the term it is meant that particles there interact with each
other via binary short-range forces), we shall here concern
ourselves with the Boltzmann gas only.

 In section 2, two plain counterexamples to
the existing theory are presented, which hints that the commonly
held notion about distribution functions is actually unsound. In
section 3, the two counterexamples are formulated with help of the
textbook methodology of collisions; it is found that only the
average distribution function over each finite velocity
solid-angle element can be calculated, while the true distribution
function is in principle not determinable. In section 4, a special
type of indeterminacy principle, or uncertainty principle, is
addressed and discussed. Section 5 concludes this paper.

\section{Counterexamples to the existing theory}
To hint that the conventional notion about distribution functions
indeed involves difficulty, two plain counterexamples against the
Boltzmann theory are addressed in this section.

The first counterexample is related to the setup shown in Fig.~1,
in which there are two dilute particle beams, produced by two
sources, colliding with each other. On the premise that the
initial distribution functions of beam 1 and beam 2 are
respectively given as
\begin{equation} f_1'({\bf r},{\bf v}_1',t)\quad{\rm and}\quad
f_2'({\bf r},{\bf v}'_2,t),
\end{equation}
we try to determine the distribution function of the beam 1
particles scattered by beam 2 particles, denoted as $f({\bf
r},{\bf v}_1,t)$. In the consideration herein, $f'_1$ and $f'_2$
are assumed to be quite general, containing no $\delta$-functions
to avoid controversy\cite{true}, which can be accomplished by
letting the two particle sources have finite temperatures and
finite outlets. For $f\equiv f({\bf r},{\bf v}_1,t)$, we write
Boltzmann's equation in the form\cite{reif}:
\begin{equation}\label{bl} \frac{\partial f}{\partial t}+
 {\bf v}_1\cdot \frac{\partial f}{\partial {\bf r}}+
\frac{{\bf F}}m \cdot \frac{\partial f}{\partial {\bf v}_1}
=\left(\frac{\delta f}{\delta t}\right)_{\rm
gain}-\left(\frac{\delta f}{\delta t}\right)_{\rm loss},
\end{equation}
where $m$ is the mass of a beam 1 particle, $\bf F$  the external
force, and $(\delta f/\delta t)_{\rm gain}$ and $(\delta f/\delta
t)_{\rm loss}$ represent the collision terms making $f$ increase
and decrease respectively.

\includegraphics*[150,582][500,720]{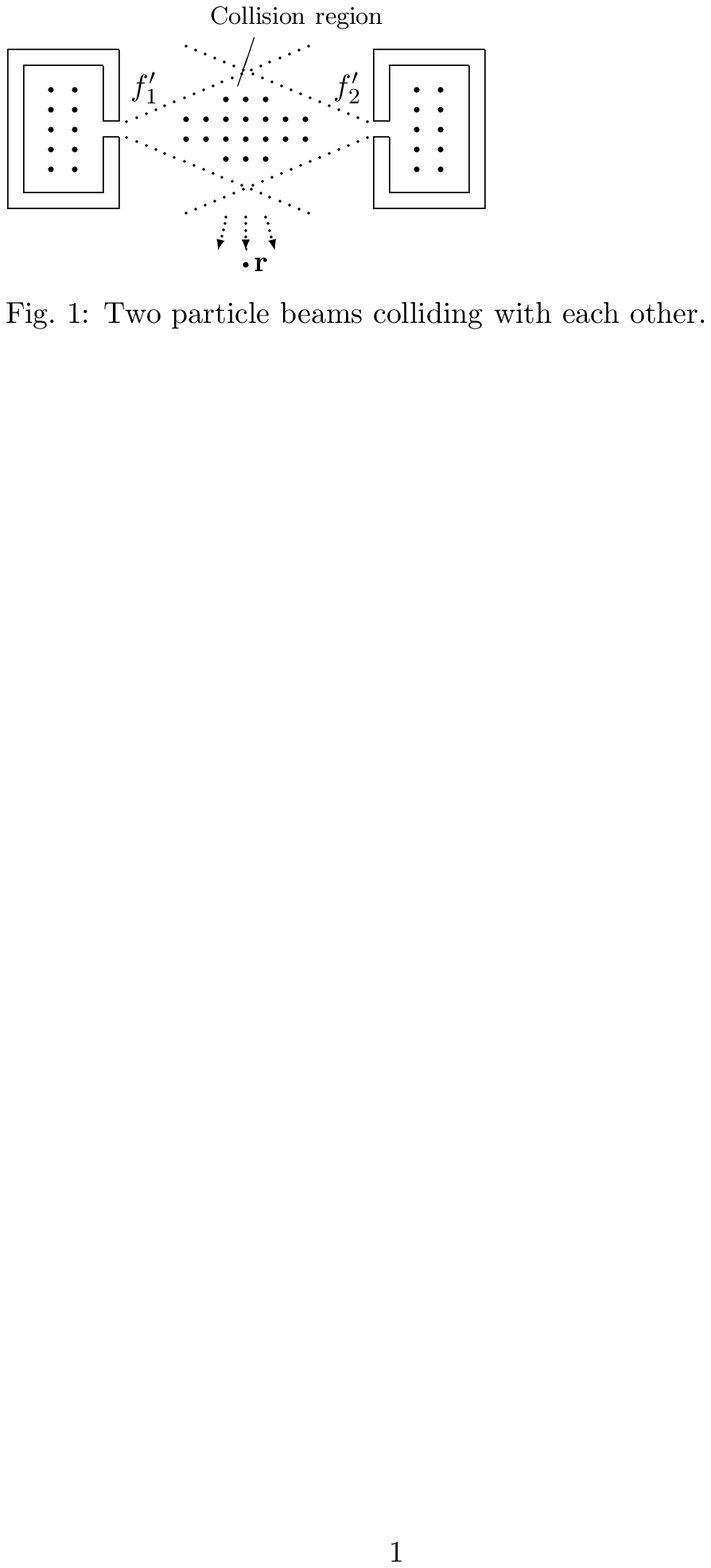}

Inside the collision region, due to the aforementioned assumption
that the two beams are dilute, $f$ must be much smaller than
$f'_1$ or $f_2'$ and thereby
\begin{equation}
\left( \frac{\delta f}{\delta t}\right)_{\rm gain}\propto f_1'f_2'
\quad{\rm and }\quad  \left(\frac{\delta f}{\delta t}\right)_{\rm
loss}\propto (ff'_1+ff'_2)\sim 0,
\end{equation}
which means that secondary collisions can be neglected. By
adopting that the situation is stationary (the sources produce
particles in a constant manner during the time of interest) and
there is no external force acting upon the particles, we arrive at
\begin{equation}\label{bl1}
{\bf v}_1\cdot \frac{\partial f}{\partial {\bf r}}
= \left(\frac{\delta f}{\delta t}\right)_{\rm gain}-
\left(\frac{\delta f}{\delta t}\right)_{\rm loss}> 0
\end{equation}
Outside the collision region, we get
\begin{equation}\label{bl2}  {\bf v}_1\cdot \frac{\partial f}{\partial {\bf r}}
=  0 .
\end{equation}
In Fig.~2, what equations (\ref{bl1}) and (\ref{bl2}) say about
the distribution function versus the distance to the system's
center is schematically depicted by the dotted curve.

\includegraphics*[150,529][500,715]{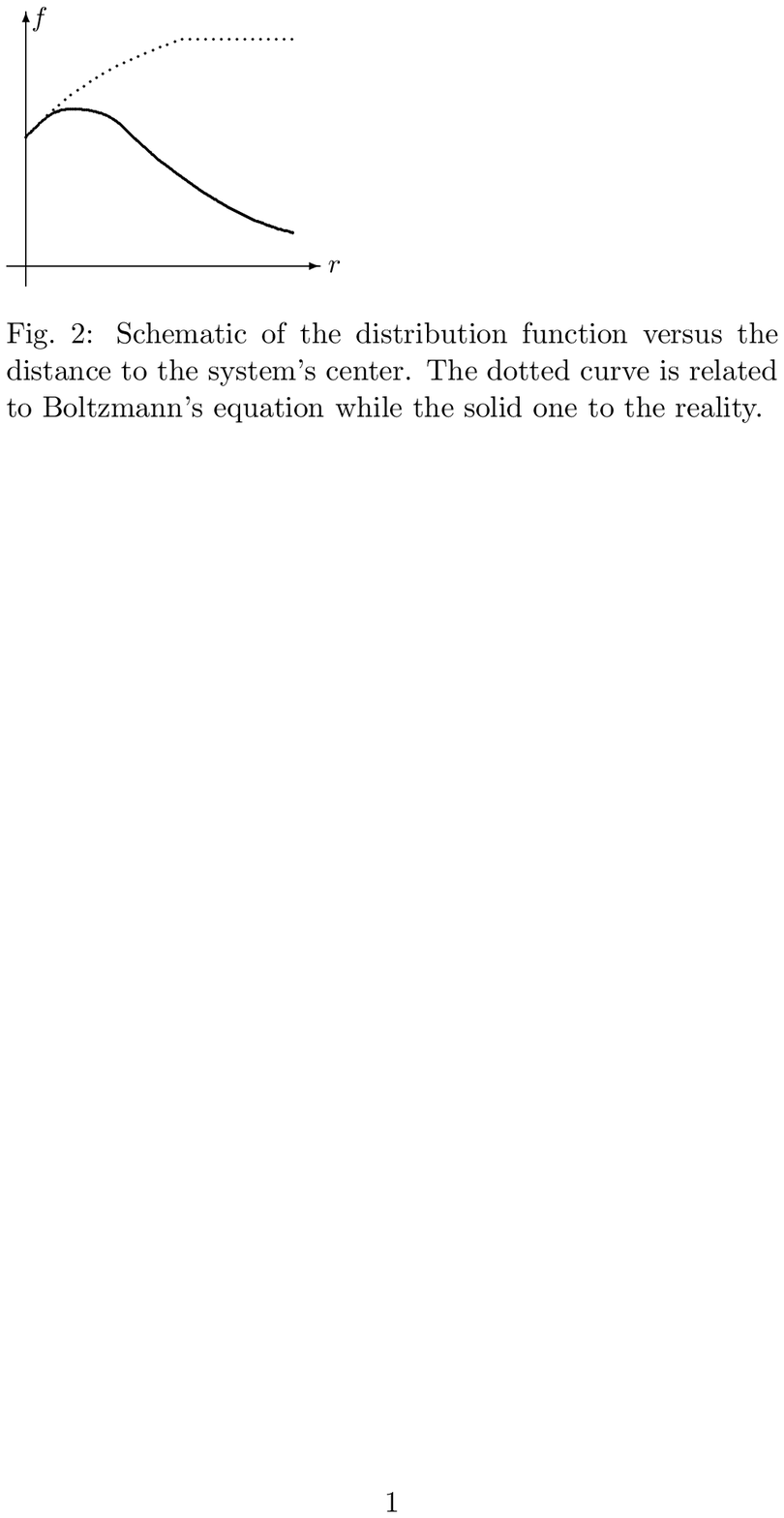}

However, our physical intuition, as well as any well-conducted
experiment, tells us something remarkably different: the
distribution function will, beyond a certain distance to the
system's center, diminish gradually to zero, as the solid curve in
Fig.~2 illustrates.

The second counterexample is shown in Fig.~3, in which the
particles of a dilute beam (of $\delta$-function type or not)
collide with a solid boundary.

\includegraphics*[150,420][500,575]{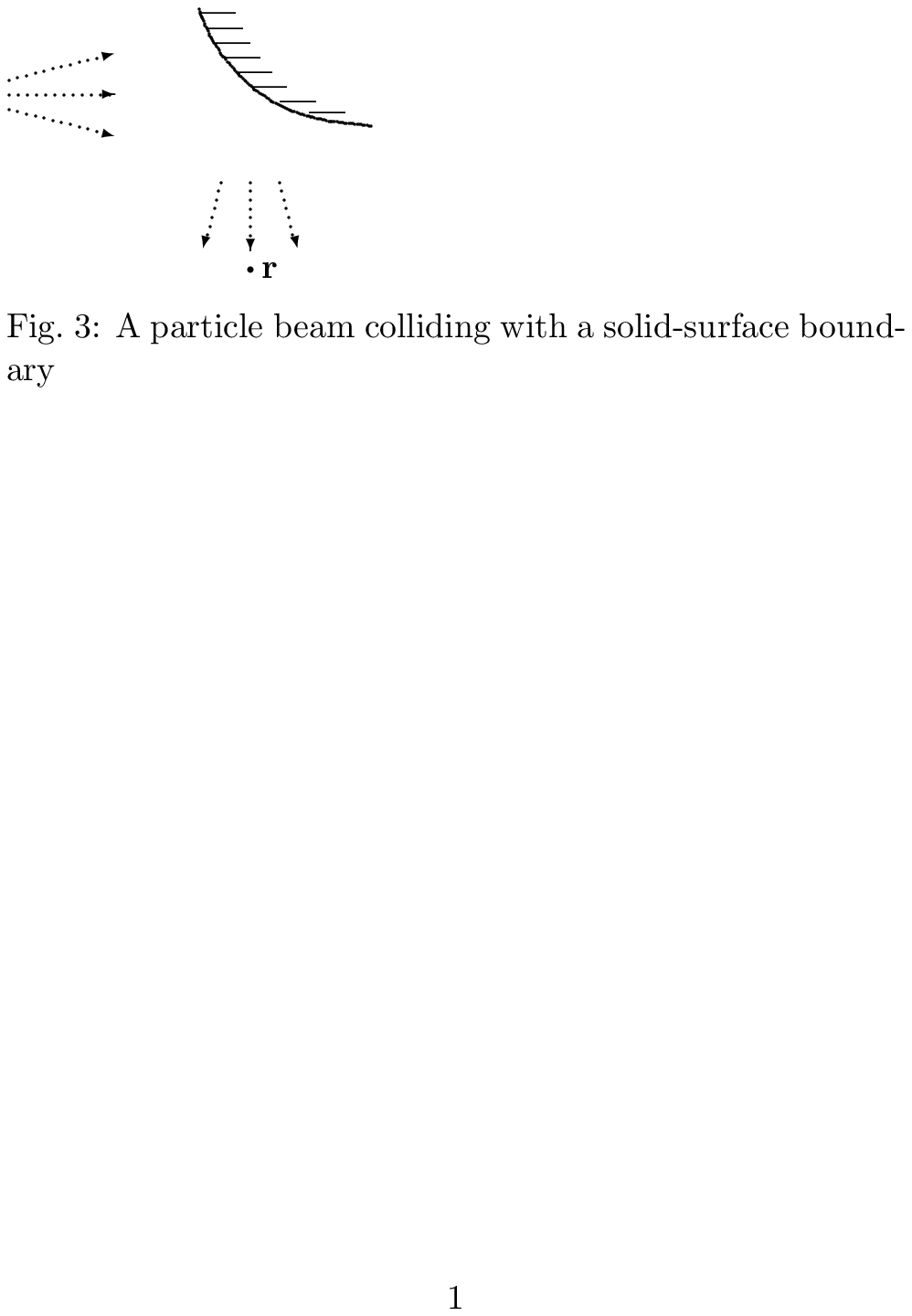}

Again, Boltzmann's equation is employed to study the distribution
function of the particles scattered by the boundary, denoted as
$f({\bf r},{\bf v}_1,t) \equiv f$. If the situation is stationary
and no external force applies, the distribution function obeys
\begin{equation} \label{bl3}
{\bf v}_1\cdot \frac{\partial f}{\partial {\bf r}}=0.
\end{equation}
In Fig.~4, what equation (\ref{bl3}) says is given by the dotted
curve while what our physical intuition, as well as real
experiments, says by the solid curve.

\includegraphics*[150,530][500,715]{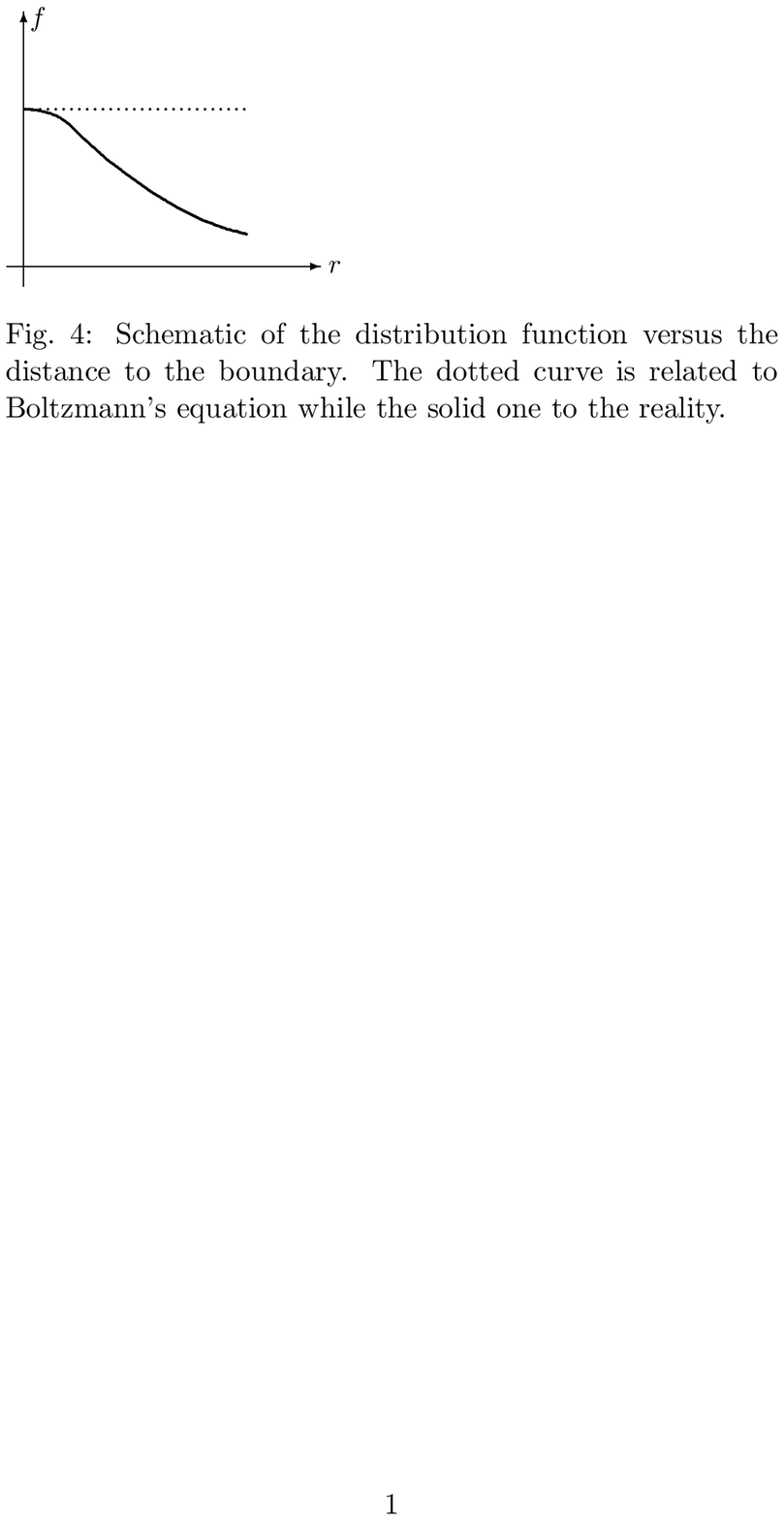}

In view of the fact that all moving particles can be deemed as
ones scattered either by other particles or by boundaries, the
foregoing two counterexamples, though somewhat heretical, possess
general significance.

\section{Formulations of the counterexamples}

In order to understand the counterexamples given in the last
section, we shall try to formulate the distribution functions by
all means available to us.

\includegraphics*[150,519][500,705]{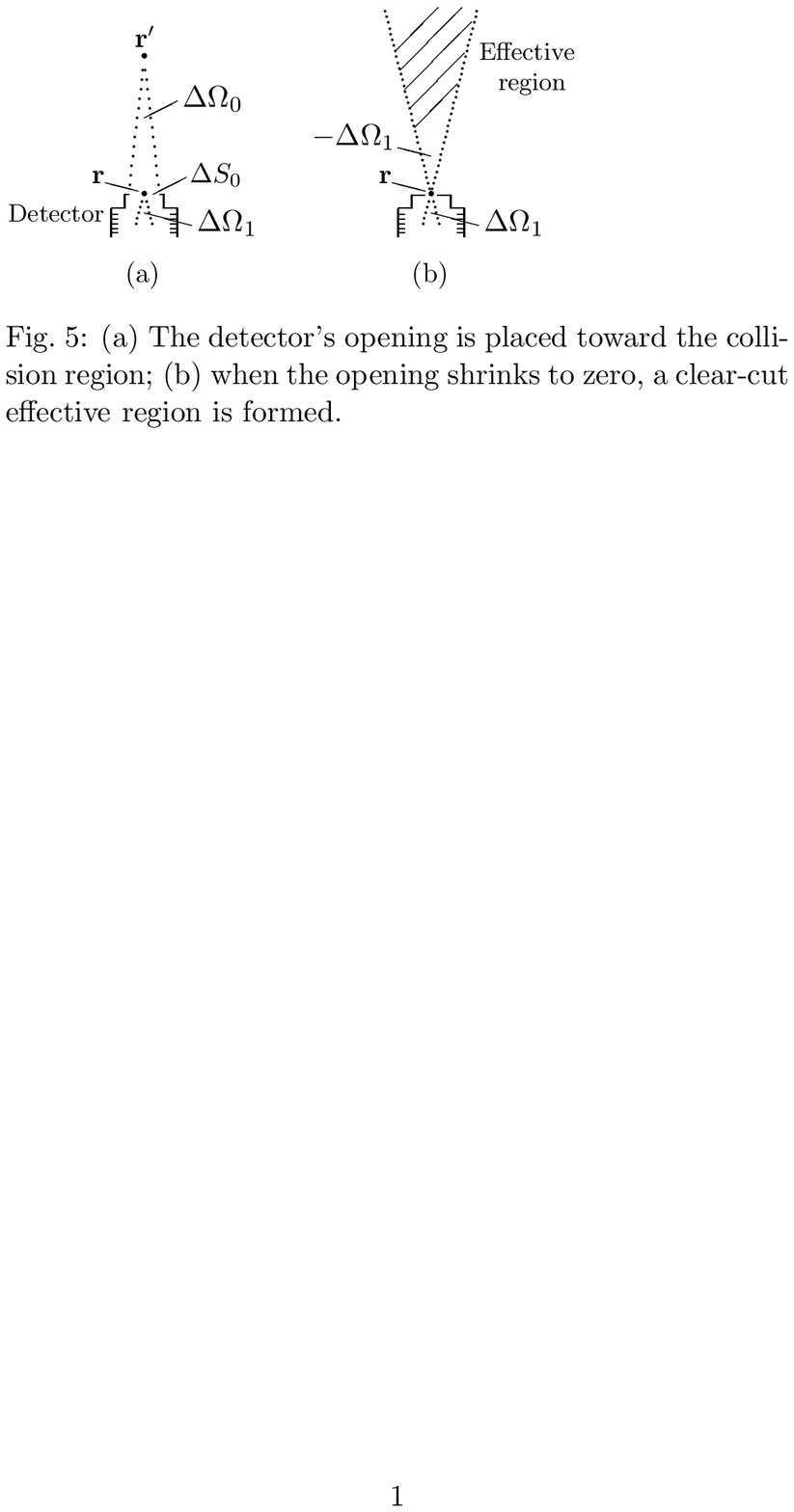}

For the first counterexample, referring to Fig.~5, a virtual
detector is placed at the point $\bf r$ where the velocity
distribution is of interest and the opening of the detector is
toward the collision region. We then assume that every beam 1
particle entering the detector within the velocity element $v_1^2
\Delta v_1\Delta\Omega_1$ ($v_1=|{\bf v}_1|$ and $\Omega_1$ is the
solid-angle of ${\bf v}_1$ in the velocity space) will be
registered as an ``effective particle''. According to the standard
theory, the distribution function is the limit of the following
ratio:
\begin{equation}\label{definition20} \frac{\Delta N_1}{\Delta{\bf
r}\cdot v_1^2 \Delta v_1 \Delta\Omega_1},
\end{equation}
in which $\Delta {\bf r}$ is a spatial volume element just inside
the detector opening and $\Delta N_1$ is the number of all the
effective particles in $\Delta {\bf r}$. It turns out that
(\ref{definition20}) can be directly and readily calculated except
that there is a compromise.

The compromise is related to the fact that (\ref{definition20})
can be evaluated only under the assumptions:
\begin{equation}\label{small}
\Delta {\bf r}\to 0 , \quad \Delta v_1\rightarrow 0 \quad{\rm
and}\quad \Delta\Omega_1 \not\to 0 .\end{equation} In what
follows, we shall do the calculation with (\ref{small}) adopted,
and later on investigate what happens if $\Delta\Omega_1$ also
shrinks to zero.

Let $\Delta S_0$ be the area of the detector opening and
$\Delta\Omega_0$ be the solid angle domain formed by $\Delta S_0$
and a point ${\bf r}'$ in the collision region, as shown in
Fig.~5(a). In order for the detector to detect the ratio
(\ref{definition20}) under (\ref{small}), $\Delta S_0$, and thus
$\Delta\Omega_0$, needs to be infinitely small. With $\Delta
\Omega_1$ being finite and $\Delta \Omega_0$ being infinitesimal,
it is found that the collisions occurring inside the region
enclosed by the spatial cone $-\Delta\Omega_1$ will possibly
produce effective particles while the collisions taking place
elsewhere will not, as shown in Fig.~5(b). For convenience, the
region, defined by ${\bf r}$ and $-\Delta\Omega_1$, is named as
the effective region and the path linking from a point in it to
the detector opening as an effective path.

Following the textbook methodology of treating collisions almost
exactly\cite{reif}, the number of all relevant effective particles
can be represented by
\begin{equation} \label{twobeams1} \Delta t' \int d{\bf r}'\int
d{\bf v}'_1 \int d{\bf v}'_2  \int d\Omega\cdot 2u \sigma(\Omega)
f_1'({\bf r}',{\bf v}'_1,t') f_2'({\bf r}',{\bf v}'_2,t') ,
\end{equation}
where $\int d{\bf r}'\cdots$ runs over the entire effective region
(over all effective paths in the sense), $2u=|{\bf v}_1'-{\bf
v}_2'|$, $\sigma$ is the cross section, $\int d\Omega\cdots$ runs
over all the passages through which effective particles move and
enter the detector, and $t'=t-{|{\bf r}-{\bf r'}|}/{v_1}$ reflects
the delay between the collision time at ${\bf r}'$ and the arrival
time at ${\bf r}$.

Since $\sigma (\Omega)$ and $d\Omega$ are conveniently defined in
the center-of-mass frame while the passages from the collision
location to the detector opening are conveniently defined in the
laboratory frame, we must do something other than what textbooks
have elaborated. With the aid of the notations ${\bf c}'\equiv
({\bf v}_1'+{\bf v}'_2)/2$, ${\bf c}\equiv ({\bf v}_1+{\bf
v}_2)/2$, ${\bf u}'\equiv ({\bf v}'_1-{\bf v}'_2)/2$, ${\bf
u}\equiv ({\bf v}_1-{\bf v}_2)/2$ and $u=|{\bf u}'|$, the
conservation laws of energy and momentum can be written as (every
particle of beam 1 and beam 2 is assumed to have the same mass)
\begin{equation} \label{conser}
{\bf c}'={\bf c}\quad {\rm and}\quad u=|{\bf u}'|=|{\bf u}|.
\end{equation}
By virtue of $u=|{\bf u}'|$, expression (\ref{twobeams1}) becomes
\begin{equation} \label{twobeams11} \Delta t' \int d{\bf r}'\int
d{\bf c}' \int d{\Omega}'\int u^2du\int d\Omega\cdot\|J\|  2u
\sigma(\Omega) f_1' f_2'
 ,
\end{equation}
where $\|J\|=\|\partial ({\bf v}_1',{\bf v}_2') /\partial ({\bf
c}',{\bf u}')\|=8$, $\int d{\Omega}'\cdots$ runs over all possible
directions of ${\bf v}_1'-{\bf v}_2'$,  $f_1'$ and $f_2'$ stand
for, respectively, \begin{equation}f_1'({\bf r}',{\bf c}'+{\bf
u}',t') \quad {\rm and} \quad  f_2'({\bf r}',{\bf c}'-{\bf u}',t')
.
\end{equation}

\includegraphics*[150,534][500,665]{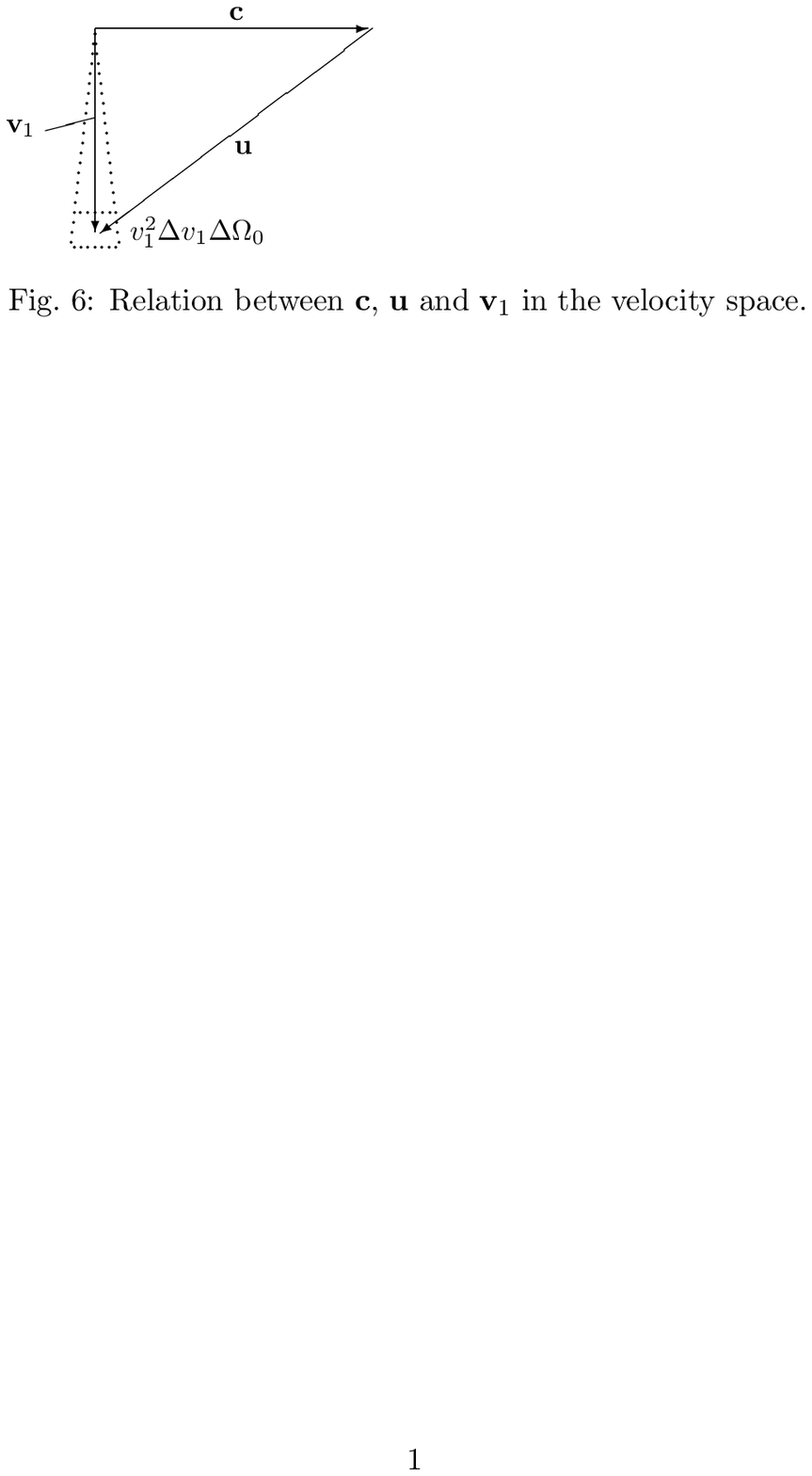}

\noindent The geometry in the velocity space, shown in Fig.~5(a)
and Fig.~6, informs us that
\begin{equation}\label{velo}
\int u^2 du\int d\Omega \cdots = v_1^2\Delta v_1 \Delta \Omega_0
\cdots. \end{equation} Fig.~5(a) also shows that the particles
will, at the detector opening, occupy the spatial volume:
\begin{equation}\label{space}
\Delta{\bf r}= |{\bf r}-{\bf r}'|^2\Delta \Omega_0 v_1\Delta
t'.\end{equation} Taking (\ref{twobeams11}), (\ref{velo}) and
(\ref{space}) into account, we can express (\ref{definition20}),
under (\ref{small}), as
\begin{equation} \label{final} f({\bf r},v_1,\Delta
\Omega_1,t)=\frac{1}{v_1 \Delta \Omega_1} \int d{\bf r}'\int d{\bf
c}' \int d{\Omega}' \frac{2 u\|J\| \sigma(\Omega) f_1' f_2'
}{|{\bf r}-{\bf r}'|^2}.
\end{equation} All quantities in (\ref{final}), explicit or
implicit, are well defined. For instance, ${\bf u}\equiv (v_1{\bf
n}_0-{\bf c}')$ with ${\bf n}_0=({\bf r}-{\bf r}')/|{\bf r}-{\bf
r}'|$, $|{\bf u}'|=|{\bf u}|=u$ while ${\bf u}'$ points in the
direction of $\Omega'$, and $\Omega$ is defined by ${\bf u}$ and
${\bf u}'$.

It is obvious that expression (\ref{final}) represents the
distribution function averaged over the given, or chosen, finite
velocity solid-angle element $\Delta\Omega_1$. At this stage,
several points should be made. (i) No approximation has been
introduced in the formulation. (ii) This formula is consistent
with the solid curve in Fig.~2; interested readers may confirm it
in real or computational experiments. (iii) The formalism lives
with discontinuity comfortably. (iv) If necessary, the secondary
collisions along each effective path can be taken into account in
a relatively easy manner\cite{chen2}.

We now look at whether or not `the true distribution function' can
be determined. As mentioned in the introduction, if the limit of
(\ref{definition20}) exists, we shall meet with it as $\Delta{\bf
r}$, $\Delta v_1$ and $\Delta\Omega_1$ tend to zero in whatever
way. That is to say, the limit of (\ref{final}), namely
\begin{equation} \label{lim} \lim\limits_{\Delta \Omega_1\to 0 }
\frac{1}{v_1 \Delta \Omega_1} \int d{\bf r}'\int d{\bf c}' \int
d{\Omega}' \frac{2 u\|J\| \sigma(\Omega) f_1' f_2' }{|{\bf r}-{\bf
r}'|^2},
\end{equation}
expresses nothing but the true distribution function. To see
whether (\ref{lim}) is mathematically meaningful, we inspect our
formulation from (\ref{definition20}) to (\ref{lim}) and examine
each of the involved limiting processes. The following three
groups of limiting processes appear very essential. (i) $\Delta
S_0\to 0$, thus $\Delta\Omega_0\to 0$, that arises from
$\Delta{\bf r} \to 0$. (ii) $\Delta \Omega_1\to 0$ that arises
from $\Delta {\bf v}_1\to 0$. (iii) The integration itself can be
deemed as a limit: the effective region is divided into many small
elements in the analysis and the final integration is performed
continuously over the region. It turns out that these limiting
processes do not get along very well. As one example, if $\Delta
\Omega_1$ becomes, when shrinking to zero, comparable to $\Delta
\Omega_0$, the very definitions of the effective region and
effective path will break down and the whole formalism will
collapse. As another example, if we let $\Delta \Omega_1$ approach
zero much slower than $\Delta \Omega_0$ does, the definition of
the effective region manages to hold and expression (\ref{lim})
will, with help of $d{\bf r}'\approx |{\bf r}-{\bf r}'|^2 \Delta
\Omega_1 dl$, become a line integral which corresponds to
Boltzmann's equation quite well. This outcome, seemingly desirable
and desired, is actually misleading. Not only that the very spirit
of the true distribution function, expressed by
(\ref{definition}), prohibits any ordering between $\Delta{\bf r}$
and $\Delta{\bf v}_1$ (one type of path-dependence), but that the
line integral yields, in terms of dealing with this example, a
curve that is almost the same as the dotted one in Fig.~2.

We now turn our attention to the second counterexample in the last
section. Interestingly, we shall encounter roughly the same
difficulty.

Based on the notion that particles scattered by boundaries behave
very much like ones emitted by boundaries\cite{kogan}, we describe
the scattered particles of this setup in terms of the emission
rate $\rho$, which is defined as such that
\begin{equation}\label{surface} d N =\rho(t',{\bf r}',v', \Omega')d t'
d S' v'^2 d v' d \Omega'\end{equation} represents the number of
the particles scattered in $d t'$ from a boundary patch $dS'$
(located at ${\bf r}'$) and emerging within the velocity range
$v'^2 dv'd\Omega'$. Generally speaking, $\rho$ depends on many
factors, such as the distribution function of the incident
particles, surface's geometry and surface's property, and $\rho$
should be determined with help of experiments. For the purpose of
this paper, we shall do our formulation on the premise that $\rho$
is completely known.

As shown in Fig.~7, a virtual detector is placed at $\bf r$;
$\Delta S_0$ and $\Delta \Omega_0$ are defined in the same way as
in the first counterexample. Furthermore, the particles entering
the detector within the velocity element $v_1^2 \Delta v_1\Delta
\Omega_1$ are also registered as effective particles and the
assumptions expressed by (\ref{small}) are also adopted at the
beginning.

\includegraphics*[150,456][500,600]{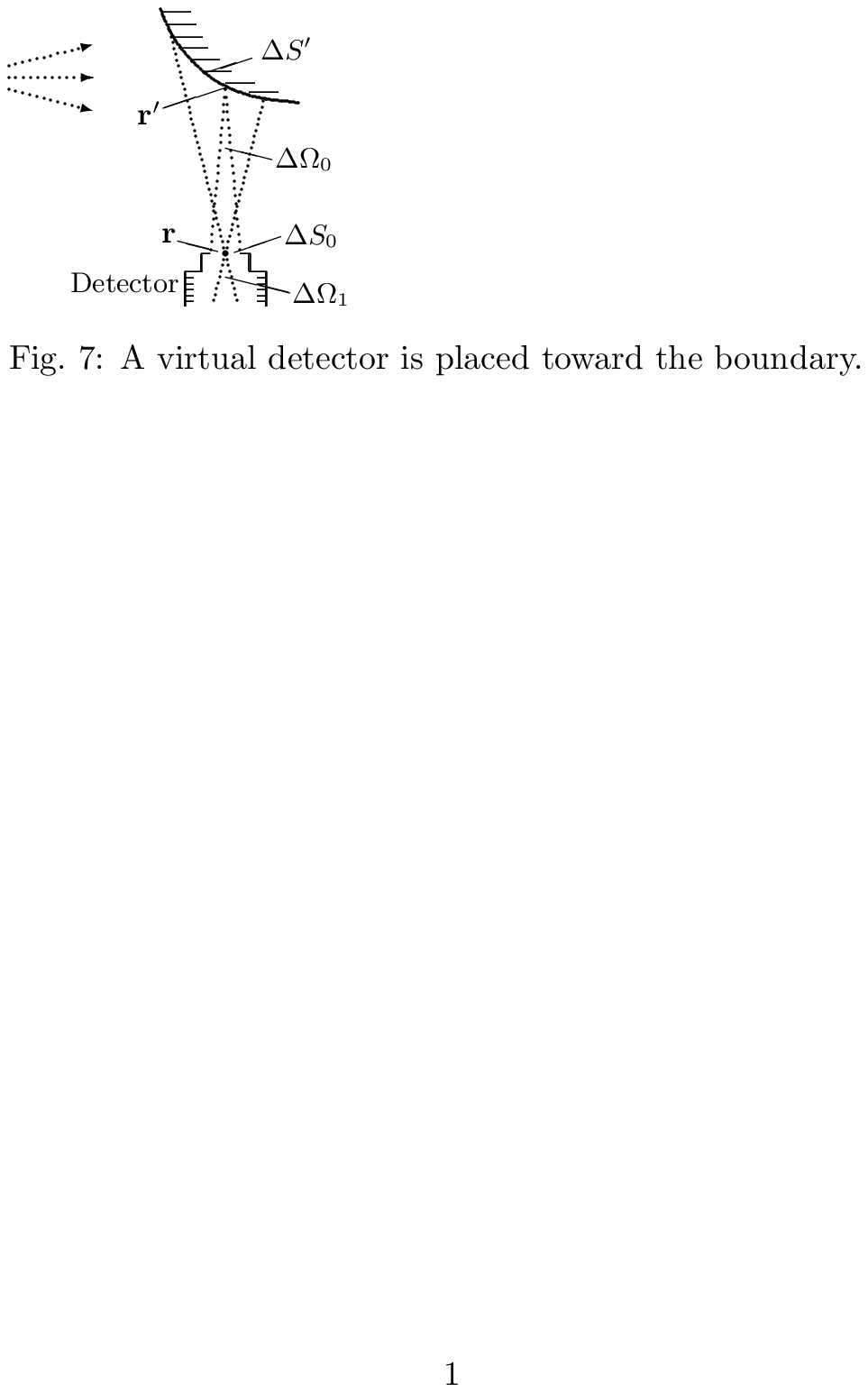}

 The area on the boundary surface enclosed by
$-\Delta\Omega_1$ (starting from $\bf r$) will be called the
effective area, denoted as $\Delta S'$. The particles emitted from
a point ${\bf r}'$ on $\Delta S'$ will, at the detector opening,
occupy the spatial volume $|{\bf r}-{\bf r}'|^2 \Delta \Omega_0 v'
dt'$. Integrating all the contributions from the effective area,
we obtain ``the distribution function'' as
\begin{equation}\label{fdelta} f(t,{\bf r},v_1, \Delta \Omega_1) =
\int_{\Delta S',\Delta v_1,\Delta \Omega_0}\frac{\rho(t',{\bf
r}',v', \Omega')d t' d S' v'^2 d v' d \Omega'}{|{\bf r}-{\bf
r}'|^2\Delta \Omega_0 v'  dt'\cdot v_1^2\Delta v_1\Delta \Omega_1}
.\end{equation}  With help of $\Delta S_0 \rightarrow 0$ (thus
$\Delta\Omega_0\rightarrow 0$) and $\Delta v_1\rightarrow 0$,
expression (\ref{fdelta}) becomes
\begin{equation}\label{df} f(t,{\bf r},v_1, \Delta \Omega_1) =
\frac{1}{v_1 \Delta \Omega_1}\int_{\Delta S'}\frac{\rho(t',{\bf
r}',v_1, \Omega') d S' }{|{\bf r}-{\bf r}'|^2 },
\end{equation}
where $t'=t-|{\bf r}-{\bf r}'|/v_1$.

Expression (\ref{df}) is nothing but the distribution function
averaged over the finite velocity solid-angle domain. Can we
arrive at the exact distribution function along the line? The
answer to it is "no" again. If $\Delta\Omega_1$ goes smaller and
becomes comparable to $\Delta \Omega_0$, the effective area will
no longer be effective and the whole formalism will no longer
hold.

Before leaving this section, it is relevant and interesting to
examine whether or not we can, for our two foregoing cases,
determine the limit:
\begin{equation}\label{definition30}
\lim\limits_{\Delta l\to 0,\Delta v_1\to 0,\Delta\Omega_1\to 0}
\frac{\Delta N_1}{\Delta S_0 \Delta l \cdot v_1^2 \Delta v_1
\Delta\Omega_1},
\end{equation}
in which $\Delta S_0$, representing the opening area of the
virtual detector, is kept to be finite (while $\Delta \Omega_1\to
0$). Expression (\ref{definition30}), though conjugate to
(\ref{final}), is not calculable in light of the fact that, to do
the evaluation, we have to divide $\Delta S_0$ into many $dS_0$
and then determine the true distribution function on each
infinitesimal $dS_0$, which cannot be done with $\Delta\Omega_1\to
0$, as just revealed.

\section{A special type of uncertainty principle}

The last section has shown that there are intrinsic and
unsurmountable difficulties to prevent us from formulating the
genuine distribution function of scattered particles (unless there
exists sure uniformness in space and in time); however, by
partitioning the velocity solid-angle space into many finite
elements, the average distribution function over each of the
elements can be calculated. For a reasonably complex system, the
taking-average strategy may be thought of as a `better-than-none'
one: once the way of partition is predetermined, how the genuine
distribution function varies within each of the finite velocity
solid-angle elements will be indeterminable and all possible
effects caused by the variation within each of them will be beyond
the investigation, which literally means that the link from any
limited number of investigations (one way of partition defines one
investigation) to the real behavior of the system is not truly
certain. The aforementioned conception, concerning in what sense
we can and cannot determine distribution functions, might be
regarded as a special type of indeterminacy principle or
uncertainty principle.

It is obvious that the integral formalism given in the last
section plays a vital role in this issue. Not only that it serves
as a mathematical demonstration on why the uncertainty principle
has to be introduced, but that it offers an effective methodology
to formulate the Boltzmann gas under the uncertainty principle.
The following perspectives may help us to compare between the
standard approach and this approach.

\begin{enumerate}

\item The standard theory is based on the tacit assumption that
all distribution functions are either differentiable or can be
approximated by differentiable functions, which is, however, in
conflict with the fact that a variety of real distribution
functions are discontinuous or quasi-discontinuous (of nonuniform
continuity). This approach, while denying the legitimacy of
differentiating distribution functions, lives with discontinuity
comfortably: the final formula is in an integral form capable of
handling discontinuity; and it yields a stepwise varying
discontinuous distribution function everywhere.

 \item As well known, one of the assumptions of
Boltzmann's equation is the time-reversibility of particle
collision, while one of the final conclusions of Botlzmann's
equation is the time-irreversibility of particle collisions; the
logic paradox there has inspired and continues to inspire a
variety of discussions in the literature. Whereas, this approach
is inherently of time-irreversibility in view of the fact that the
resultant distribution function is obtained by integrating the
distribution functions at previous times over a region (or area),
via which a lot of historical and detailed information is erased
explicitly.

\item In the standard approach, after a distribution function is
obtained by solving Boltzmann's equation, presumably with help of
specific initial and boundary conditions, the task of the kinetic
theory is finished in a once-for-all manner. In this approach, we
may alter the partition of finite velocity solid-angle elements
and restart the calculation. By repeatedly doing so, a number of
resultant distribution functions for the system can be obtained.
If the system is sufficiently complex and the evolution time is
sufficiently long, it is quite possible to find that same initial
and boundary conditions result in dramatically different outcomes.
This is partially related to the uncertainty nature of this
approach and partially related to the non-uniqueness nature of our
world.

\item As one of the most important features, this approach
produces, in many simple cases, definite results that are directly
verifiable with help of experimental or computational means. In
particular, the two counterexamples against the standard theory
presented in Sect.~2 have been nicely treated in Sect.~3.

\end{enumerate}

Many issues, closely related and not so closely related, have yet
to be explored; some of them are discussed by the author
elsewhere\cite{chen3}.

\section{Conclusion}

It is often said that to decide whether a theory bears physical
truth, one must be able to falsify it. Unfortunately, Boltzmann's
theory has been established for so long that people in this
community lose interest in doing so. This paper should, at least,
serve as a clue that a considerable number of concepts and
methodologies of the existing statistical theory need serious
reconsideration.

\section*{Acknowledgement}
Communication with Prof. Oliver Penrose is gratefully
acknowledged. The author also thanks professors Hanying Guo, Ke Wu
and Keying Guan for helpful discussion.

\end{document}